\documentstyle[11pt]{article}
\title{
	\begin{flushright}
	{\normalsize TPI-MINN-95/21-T\\
	NUC-MINN-95/18-T \\
	UMN-TH-1401/95\\
       CERN-TH/95-168\\
	June 1995\\}
	\end{flushright}
\vskip1cm
\bf The Return of the Prodigal Goldstone Boson}
\author{
	J. Kapusta$^a$, D. Kharzeev$^{b,c}$ and L. McLerran$^a$\\
	{\small\it $^a$ School of Physics and Astronomy,
	University of Minnesota, Minneapolis, MN 55455} \\
	{\small\it $^b$ Theory Division, CERN,
	Geneva, Switzerland}\\
        {\small\it $^c$ Fakult\"at f\"ur Physik, Universit\"at Bielefeld,
        Bielefeld, Germany } \\
 }

\date{}

\begin{document}

\maketitle

\begin{center}
{\bf Abstract}\\
\end{center}
We propose that the mass of the $\eta^\prime$ meson is a particularly sensitive
probe of the properties of finite energy density hadronic matter and
quark gluon plasma.  We argue that the mass of the $\eta^\prime$
excitation in hot and dense matter should be small, and therefore
that the $\eta^\prime$ production cross section should be much increased
relative to that for pp collisions. This may have observable
consequences in dilepton and diphoton experiments.

\vskip1.5cm
\vfill \eject

\section{Introduction}

One of the great mysteries in the quark model was why there is no
ninth Goldstone boson whose mass is somewhere between that of
the pion and the kaon.  Roughly stated, the problem
is that in the limit of massless quarks, the quark model has
a $U(3)$ chiral symmetry.  This chiral symmetry, when broken,
predicts the existence of 9 massless Goldstone bosons. In nature,
however, there are only 8 light mesons.

The problem is resolved by the Adler-Bell-Jackiw $U(1)$ anomaly
\cite{adler}-\cite{jackiw}:
the $U(1)$ part of the $U(3)$ symmetry is explicitly broken
by interactions.  It is possible to show explicitly
that instantons \cite{polyakov,thooft} dynamically induce the $U(1)$
chiral symmetry breaking.  This results in giving the
ninth Goldstone boson a mass.  The would--be ninth Goldstone boson is
presumably
the $\eta^\prime$, which has a mass of nearly one GeV.

As the density of matter is increased, it is expected that the
effects induced by the tunneling between different topological vacua of QCD
will rapidly disappear \cite{gpy}-\cite{DM}.
Let us briefly recall the origin of this belief, based on the example of
the instanton solution realizing this tunneling.
The amplitude ${\cal T}$ of the tunneling transition,
calculated in the quasiclassical approximation of instanton
configurations, is
\begin{equation}
{\cal T}_{\rm instanton} \sim e^{-S_E} \sim e^{-2\pi /\alpha_S}\, ,
\label{dense}
\end{equation}
where $S_E$ is the Euclidean action of the instanton solution.
It is expected that the effects of finite energy density will make
$\alpha_S$ density dependent such that for large energy densities
\begin{equation}
\alpha_S \sim {{24\pi} \over {(33-2N_f)\ln(\epsilon/\Lambda^4)}} \, ,
\label{alpha}
\end{equation}
where $\epsilon$ is the energy density and $\Lambda \sim 200$ MeV.
As the energy density increases, the effects of instantons
rapidly decrease.  Note that $\Lambda^4 \sim 200$ MeV/fm$^3$ is a
relatively low energy density.

We therefore expect that as the energy density of hadronic matter is increased,
the mass of the $\eta^\prime$ will be a rapidly falling function of energy
density.  In the quark gluon plasma, we expect that excitations
with the quantum numbers of the $\eta^\prime$ will become almost
mass--degenerate, modulo current quark mass corrections, with
excitations with quantum numbers of the octet of pseudoscalar
Goldstone bosons.
This is manifest in the quark model since there will be no
penalty for making an isospin singlet configuration
of quarks relative to an isotriplet configuration.

The plan of this paper is as follows. In Section 2 we recall
the mechanism responsible for the large mass of the $\eta'$ in QCD, and
argue about the properties of the $\eta'$ at high densities.
In Section 3 we discuss the dynamics of $\eta'$ production
and annihilation in hot and dense matter. In Section 4 we discuss several
expected signals of the proposed enhancement of $\eta'$ production in
dense matter and claim possible experimental evidence in favor
of our scenario.

\section{Axial anomaly, ghost, and $\eta'$ at high densities}

Consider a quark--antiquark pseudoscalar flavor--singlet field
\begin{equation}
|\eta_0 \rangle = {1 \over{ \sqrt{3}}}
|\bar{u}u + \bar{d}d + \bar{s}s \rangle \, . \label{eta}
\end{equation}
The divergence of the corresponding flavor--singlet axial current
acquires an anomalous part, due to the interaction with gluon fields,
which does not disappear in the chiral limit
$m \to 0$ of massless quarks:
\begin{equation}
\partial^{\mu} J_{5 \mu}^0 = 2i\sum_f m_f \bar{q_f}\gamma_5 q_f +
2N_f {g^2 \over {16\pi^2}} {\rm Tr}\left(G_{\mu\nu}\tilde{G}^{\mu\nu}
\right) \, .
\end{equation}
This anomalous part may be written as the full divergence
of the gauge--dependent topological current
\begin{equation}
K_{\mu}=2N_f {g^2 \over{16\pi^2}} \epsilon_{\mu\nu\lambda\rho}
{\rm Tr}\left(G^{\nu\lambda}A^{\rho}\right) \, , \label{topcur}
\end{equation}
so that in the chiral limit one has the Adler--Bardeen relation
\begin{equation}
\partial^{\mu} J_{5 \mu}^0 = \partial^{\mu} K_{\mu} \, . \label{AB}
\end{equation}
It is possible to introduce a new axial current
\begin{equation}
J_{5 \mu} = J_{5 \mu}^0 - K_{\mu} \, ,
\end{equation}
which is explicitly conserved in the chiral limit.
\begin{equation}
\partial^{\mu} J_{5 \mu} = 2i\sum_f m_f \bar{q_f}\gamma_5 q_f
\end{equation}
The corresponding charge
\begin{equation}
Q_5 = \int d^3x \, J_{5 0} \, , \label{charge}
\end{equation}
is naively expected to be conserved.
Since this charge is the generator of the $U(1)_A$ symmetry,
and this symmetry is not observed in the hadron spectrum (no
parity doublets exist), we have to presume that the symmetry
is spontaneously broken. This would lead to the appearance
of a nearly massless Goldstone boson field (\ref{eta}).
In nature, however, the physical $\eta'$ meson has a large mass
of almost one GeV and therefore cannot be considered a Goldstone
boson.

To check if the charge (\ref{charge}) is really conserved,
one can integrate the divergence of the current $J_{5 \mu}$
over Euclidean 4--space. After the spatial integration is performed,
the result can be represented as
\begin{equation}
\int_{-\infty}^{+\infty} dt {d Q_5 \over{d t}} = 2 N_f \nu [G] \, ,
\end{equation}
where
\begin{equation}
\nu [G] = 2N_f {g^2 \over {32\pi^2}}
\int d^4 x\ {\rm Tr}\left(G_{\mu\nu}\tilde{G}^{\mu\nu}\right) \label{top}
\end{equation}
is the so--called topological charge. It is equal to zero in
Abelian theories, but in QCD $\nu [G]\neq 0$: the one--instanton solution,
for example, yields $\nu [G]=1$. Therefore the charge (\ref{charge})
is not a conserved quantity, and going from (Euclidean) $t=-\infty$
to $t=+\infty$ it changes by
\begin{equation}
\Delta Q_5 = 2 N_f \nu [G]\, .
\end{equation}
Therefore the existence of non--trivial topological solutions
explicitly breaks the $U(1)_A$ symmetry, resulting in the
vanishing of the corresponding Goldstone mode.

As we have already mentioned in the Introduction, the instanton density
vanishes in the high energy density limit as $g^2=4\pi\alpha_S \to 0$.
We therefore expect that in dense matter the ensemble averaged
axial charge $Q_5$ will be conserved.
\begin{equation}
\frac{d \langle Q_5 \rangle }{dt} = 0
\end{equation}
If the $U(1)_A$ symmetry is still spontaneously broken at
very high densities, it would imply the return of the ninth Goldstone
boson!

Even though the arguments presented above explain on a qualitative
level why the physical $\eta'$ is not a  Goldstone excitation, and under
what circumstances can it again become one, it is
instructive for our purposes to establish the actual relation between the
properties of the vacuum and the mass of the $\eta'$.
To do this we follow the approach developed by Witten \cite{Witten} and
Veneziano \cite{Veneziano}.  They noticed that the non--vanishing of the
topological charge (\ref{top}) implies the existence of an unphysical
massless pole, introduced earlier by Kogut and Susskind \cite{KS}, in
the correlator of the topological current (\ref{topcur}).
Such a pole means the existence
of a massless excitation, or ``ghost", which should reflect some
fundamental symmetry of the theory. As was shown by Dyakonov and Eides
\cite{DE}, the origin of this excitation in QCD is the
periodicity of the potential energy of the vacuum with respect to the
collective coordinate
\begin{equation}
X = \int d^3x \, K_0(x) \, .
\end{equation}
The potential barriers separating different vacua are penetrable,
by instantons for example, and
the massless ghost just corresponds to this degree of freedom in the
theory. If one introduces the propagator
$\langle a_{\alpha}a_{\beta} \rangle$ of the
ghost field $a_{\alpha}$,
the residue of the ghost contribution $\lambda$ can be defined as
\begin{equation}
\langle 0|K_{\alpha}K_{\beta}|0 \rangle = \lambda^4 \,
\langle 0|a_{\alpha}a_{\beta}|0 \rangle \, ,
\end{equation}
so that as $q^2 \to 0$
\begin{equation}
q_{\alpha}q_{\beta} \langle 0|K_{\alpha}K_{\beta}|0 \rangle =
\langle 0|\nu \nu|0 \rangle = - \lambda^4
\neq 0. \label{res}
\end{equation}
Note that, apart from the ghost contribution, the propagator of the
topological current also contains the normal gluon part.

The field (\ref{eta}) can now mix with the ghost, the amplitude of
mixing being of order $\lambda^2/f_{\eta'}$, where  $f_{\eta'}$ is the
$\eta'$ decay constant.  As a result of this mixing the physical $\eta'$
acquires an additional mass
\begin{equation}
\Delta m \simeq {\lambda^2/f_{\eta'}} \, , \label{mas1}
\end{equation}
so that the mass of the $\eta'$ does not vanish in the chiral
limit.
\begin{equation}
m_{\eta'}^2 = m_0^2 + (\Delta m)^2 \label{mas2}
\end{equation}
The mass of the bare $\eta'$ field (\ref{eta}) can simply be estimated
in the free quark model as
\begin{equation}
m_0^2 = {1 \over 3} ( 2 m_{K}^2 + m_{\pi}^2 ) \, . \label{mass}
\end{equation}
At high energy densities we expect that the density of instantons
will diminish, the ghosts will disappear, and the $\eta'$
will be (almost) entirely described by the field (\ref{eta}), whose
mass will then be given by (\ref{mass}) and equal to $m_0\simeq 400$ MeV.

Of course in nature the situation is likely to be a bit more complicated.
Indeed, the mass eigenstates in the isosinglet channel are not the
$\eta$ and $\eta'$, but the nonstrange and strange states
$|\eta_{\rm NS}\rangle = |\bar{u}u + \bar{d}d\rangle/\sqrt{2}$ and
$|\eta_{\rm S}\rangle = |\bar{s}s\rangle$. These states can only mix
if one allows for intermediate gluon states. The extreme assumption
that the only allowed gluonic states are
non-perturbative ghost-like states would lead to the conclusion that
at high densities, when ghosts disappear, the physical isosinglet excitations
will be $\eta_{\rm NS}$ and $\eta_{\rm S}$. Their masses will then be
$m_{\rm NS}^2=m_{\pi}^2$ and
$m_{\rm S}^2=2m_K^2 - m_{\pi}^2$; $m_{\rm S} \simeq 700$ MeV.
However, normal gluonic states
certainly contribute, and we expect that the states $\eta_{\rm NS}$
and $\eta_{\rm S}$ will mix
even at high densities, even though this mixing will probably not yield
the states with the $\eta$ and $\eta'$ quark wave functions.
We expect also that as a consequence of the effects discussed above
the $\eta-\eta'$ mixing will be strongly dependent
on energy density, and the physical $\eta$ mass will decrease too.
Nevertheless, since the topological and perturbative gluonic effects are
very difficult to separate, for the sake of argument we will assume in the
rest of this paper that the $\eta'$ quark content at any density is given by
(\ref{eta}).

\section{Dynamics}

Production cross sections for light mesons are typically of
the order predicted by the Hagedorn model,
\begin{equation}
\sigma_i \sim g_i (M/2\pi)^{3/2}\, {\rm e}^{-M/T_H} \, ,
\end{equation}
when the particle mass is large compared to $T_H \sim 160$ MeV.  The
quantity $g_i$ is the number of internal degrees of
freedom of the i'th particle species.  For pions this same model gives
\begin{equation}
\sigma_\pi \sim g_i/\pi^2 \, .
\end{equation}
Using this rather simple model we see that the expected cross section of
$\eta'$ production is quite small, $\sigma_{\eta^\prime}/\sigma_{\pi^0}
\sim 2 \times 10^{-2}$.

Now suppose that the $\eta^\prime$ is made in a dense environment.
Here we expect that the mass of the $\eta^\prime$ is small, and the
particle ratio
$N_{\eta^\prime} /N_{\pi^0} \sim 1$.  If the $\eta^\prime$ becomes
a Goldstone boson we might get a factor of up to 50
enhancement in the production cross section!
This should of course be considered only as an absolute upper bound
for the enhancement; the strange quark mass effects (see (\ref{mass}))
result in a more moderate enhancement factor of 16, and if the  $\eta'$ at
high densities becomes an $|\bar{s}s\rangle$ state according to the
scenario described at the end of the previous section the enhancement factor
will be equal to a relatively modest value of 3.

After an $\eta^\prime$ is produced it must survive subsequent hadronic
interactions until it has escaped the matter.  The
$\eta^\prime$ lifetime in vacuum is about 1000 fm/c; if there were
no interactions with surrounding particles, it would certainly survive
the time it takes for the hadronic matter produced
in heavy ion collisions to dissipate.

It is amazing that the results presented in the previous Section imply
that the $\eta'$ should decouple from high density matter and therefore
most likely cannot be absorbed. To see this, we will follow the line
of reasoning developed in refs. \cite{Ven} and \cite{Efr}.

Let us first note that the Adler--Bardeen relation (\ref{AB}), and an
analog of the PCAC for the $\eta'$ field,
\begin{equation}
\eta'(x) = {1 \over {m_{\eta'}^2 f_{\eta'}^2}} \partial^{\mu}J^0_{5\mu}
\, ,
\end{equation}
suggest the existence of a relation between the matrix elements of the
$\eta'$ field and of the topological charge (\ref{top}).
With this in mind, we consider a nonsymmetric matrix element of the
topological current (\ref{topcur}) between some hadronic
states\footnote{In principle, one can consider the matrix elements taken
over the ensemble as a whole.}. For definiteness we consider
nucleons explicitly here.  It has the following general form:
\begin{equation}
\langle p'|K_{\nu}|p \rangle = \bar{u}(p')[\gamma_{\nu}\gamma_5
G_1(q^2) + q_{\nu}\gamma_5 G_2(q^2)] u(p) \, ,
\end{equation}
where $q=p-p'$, $\bar{u}, u$ are the nucleon wave functions, and $G_{1,2}$
are the form factors.
Consider the matrix element $\langle 0|\partial^{\nu}K_{\nu}
|\bar{N}N \rangle$ in the
cross channel. Saturating it by the $\eta'$ pole, one obtains
\begin{equation}
q^2 G_2(q^2) = \langle 0|\nu|\eta' \rangle {1 \over {q^2 - m_{\eta'}^2}}
\langle \eta'|\bar{N}N \rangle
\, , \label{form}
\end{equation}
where the last matrix element is just the $\eta'$ coupling constant
$g_{\eta' NN}$.
The first matrix element can be evaluated
by using the Lehmann--Symanzik--Zimmerman reduction formula in the following
form:
\begin{eqnarray}
\langle 0|\nu|\eta'\rangle &=& \int d^4x {\rm e}^{iq \cdot x}
(- \partial^4_x + m^2)
\langle 0|T\{\nu \eta'(x)\}|0 \rangle \nonumber \\
&=& - {-q^2 + m_{\eta'}^2 \over{m_{\eta'}^2 f^2}}\
\langle 0|T\{\nu\nu\}|0 \rangle \, .
\label{red}
\end{eqnarray}
As $q^2\to 0$ we get, from (\ref{form}), (\ref{red}) and (\ref{res}), that
\begin{equation}
q^2 G_2(q^2) \sim \frac{\lambda^4 g_{\eta'NN}}{m^2_{\eta'}f_{\eta'}} =
f_{\eta'} g_{\eta'NN},
\label{fin}
\end{equation}
where at the last step we used the relation $m_{\eta'}\simeq
\lambda^2/f_{\eta'}$,
valid in the chiral limit (see (\ref{mas1}), (\ref{mas2})).

In the absence of ghosts, which we expect is the case in high density
matter, the form factor $G_2(q^2)$ does not possess a zero--mass pole,
and the l.h.s.
of (\ref{fin}) is equal to zero at $q^2 = 0$. Therefore, since
$f_{\eta'} \neq 0$,
we are led to the conclusion that at high densities the coupling of the
$\eta'$ vanishes and it decouples from (nonGoldstonic) matter.
A parallel discussion for the coupling of an $\eta'$ with two
$\rho$ mesons \cite{rho} can be given with a similar conclusion.

Next, consider moderate to low energy density matter where pions are the
most abundant constituents.  Then we need to know the cross
section for the annihilation reaction
$\pi^+ + \eta' \rightarrow \pi^+ + \rho^0$, which is exothermic, and
the isospin-related cross sections.  The rate can be calculated in the
low temperature limit using a low energy effective Lagrangian.

The cross section for $\pi(p_1) + \eta'(p_2) \rightarrow
\pi(p_1') + \rho(p_2')$ is dominated
by the exchange of a $\rho$-meson in the $t$-channel.  The $\rho\pi\pi$
vertex is well-known, and the $\eta'\rho\rho$ vertex is the anomalous
one \cite{Gomm,Durso}.  The matrix element is
\begin{equation}
{\cal M} = g_{\eta'\rho\rho} \, p_{2 \alpha} \, p'_{2 \beta} \,
\epsilon^{\alpha \beta \nu \tau}
\left[ - \frac{g_{\mu\nu}}{q^2-m_{\rho}^2} + \frac{q_{\mu}q_{\nu}}
{(q^2-m_{\rho}^2)m_{\rho}^2} \right] g_{\rho\pi\pi} \, (p_1+p'_1)^{\mu}
\, \varepsilon_{\tau}(p_2') \, ,
\end{equation}
where $q = p'_1 - p_1$.  The total cross section for one charge
configuration works out to be
\begin{displaymath}
\sigma_0(s) = \frac{g_{\rho\pi\pi}^2 g_{\eta'\rho\rho}^2}{16\pi
p_{cm}^2} \left\{\left(t_+ - t_- \right)
+ \left(t_+ + t_- -2m_{\rho}^2 \right)
\ln\left( \frac{m_{\rho}^2 - t_-}{m_{\rho}^2 - t_+} \right) + \right.
\end{displaymath}
\begin{equation}
\left. \frac{(t_+ - t_-)}{(m_{\rho}^2 - t_-)(m_{\rho}^2 - t_+)}
\left[ - m_{\rho}^2 (t_+ + t_-) + m_{\rho}^4
+m_{\pi}^2 (m_{\eta'}^2 - m_{\rho}^2)^2/s \right] \right\} \, .
\end{equation}
Here $t_+$ and $t_-$ are the kinematic limits of $t$.

{}From the decay rate for $\rho \rightarrow \pi\pi$ we know that
$g_{\rho\pi\pi}^2/4\pi$ = 2.90.  From the decay rate for
$\eta' \rightarrow \rho \gamma$ \cite{pbook}, together with vector meson
dominance \cite{Gomm,Durso}, we get $g_{\eta'\rho\rho}$ = 3.96 $\times$
10$^{-3}$/MeV or, more usefully, $g_{\eta'\rho\rho}^2$ = 6.10 mb.
It may be noted that this value is consistent with that predicted by
gauging the Wess-Zumino term, which is
\begin{displaymath}
g_{\eta'\rho\rho} =
\frac{g_{\rho\pi\pi}^2}{16 \pi^2 f_{\pi}} \left(\sqrt{6} \cos\theta_P
+ \sqrt{3} \sin\theta_P \right) \, ,
\end{displaymath}
where $\theta_P$ is a
pseudoscalar mixing angle with a value of about $-20 \pm 5$ degrees
\cite{Gomm,Durso,standard}.

The annihilation cross section vanishes at threshold and rises
monotonically with $s$.  Although thermal averaging can be done
numerically to obtain the rate, we shall be content with the following simple
estimate.  For a collision between an $\eta'$ and a pion the average
value of $s$ at temperature $T$ is easily found to be
\begin{displaymath}
\langle s \rangle = (m_{\eta'}+m_{\pi})^2 + 6 m_{\eta'} T \, .
\end{displaymath}
At $T$ = 150 MeV, $\sqrt{ \langle s \rangle}$ = 1.44 GeV.  At this value,
$\sigma_0$ = 2.6 mb.  The mean free path $l$ for $\eta'$ annihilation
is estimated from
\begin{equation}
l^{-1} = \sum_{ij} \sigma_{ij} n_i = 2 \sigma_0 n \, ,
\end{equation}
where the sum is over all channels, $n$ is the total pion number
density, and $\sigma_0$ is evaluated at the average $\sqrt{s}$.
For temperatures comparable to or greater than the pion mass
the number density is approximately 0.365 $T^3$.  At $T$ = 150 MeV
the mean free path for annihilation is 12 fm.  It gets even bigger
as the temperature decreases.  Since the $\eta'$ decouples near the
phase transition temperature,
where the present estimate is not valid, we may conclude that
$\eta'$s will not annihilate to any appreciable degree
at any temperature during the expansion.

It might seem paradoxical to argue that the $\eta'$ decouples
at high density yet is produced in roughly equal abundance with
the pion.  Actually, there is no paradox.  Suppose that quark
gluon plasma is formed initially.  When it hadronizes, all
Goldstone bosons will be produced in roughly equal numbers
by condensation of the quark and gluon fields.  Suppose that
high density hadronic matter is formed initially, not quark
gluon plasma.  Then the initial state is formed via meson
production in elementary nucleon-nucleon collisions.  Many pions
will be produced.  In this environment, the $\eta'$ mass
will be low.  Since there is no suppression of transitions among
the Goldstone bosons themselves, the $\eta'$ mesons will come to, or
at least approach, chemical equilibrium with pions, kaons and
$\eta$ mesons.

\section{Signals}

During the expansion and cooling phase, the $\eta^\prime$
propagates in the background field of the surrounding hadronic matter.
This background
field increases the $\eta^\prime$ mass as the hadronic matter becomes more
dilute.  Due to energy conservation, any motion of the $\eta^\prime$ relative
to this medium will be damped, and the $\eta^\prime$ will come to rest.
As a consequence, the $\eta^\prime$ will be strongly coupled to
any collective flow of matter, and the $p_T$ distribution of
$\eta^\prime$ may be strongly distorted relative
to that in pp collisions.

When the matter is at high energy density there will be mixing
between the collective excitations which will become the
$\eta$ and $\eta'$ in the vacuum, so an enhancement of the
$\eta'$ will lead to an enhancement of the $\eta$ too.
In addition, an important decay mode of the vacuum $\eta^\prime$ is
into $\eta$ with a branching ratio of 65\%, leading to an enhancement
of $\eta$ after the breakup of hadronic matter occurs.

There are several places where one might see the effects of the return
of the ninth Goldstone boson.  First, one might study low mass dileptons
in the region above the $\pi^0$ Dalitz pairs and below the $\rho$.
If the $\eta^\prime /\pi^0$ ratio is enhanced,
there would be an enhancement due to the $\eta^\prime \rightarrow
e^+e^- \gamma$ decay mode.  In Figure 1 we display the data as
measured in the CERES experiment \cite{ceres}; the paucity of dileptons
in the mass region between the $\pi^0$ and the $\rho$ was also seen by the
HELIOS experiment \cite{helios}. The contributions from measured and
assumed abundances of $\pi^0$, $\eta$, $\rho$, $\omega$, $\eta'$ and
$\phi$ are shown explicitly taking into account the acceptance and
resolution of the detector.  In Figure 2
we have scaled the computed $\eta^\prime$ contribution by 50 and 16,
corresponding to the ratios $\eta^\prime /\pi^0 = 1$ and $0.3$, where the
latter value arises from taking into account the strange quark mass effects
 - see (\ref{mass}).  To these were added the contributions from
the other mesons.  With the enhancement factor of 50 the
result is a little too big in the region between 50 and 250 MeV,
exceeding two data points by about two standard deviations.  Otherwise
the representation of the data is very good.  With the enhancement factor
of 16 there is also a good representation, although the curve
consistently falls below the data points by about one standard
deviation between 350 and 850 MeV.  We have made no attempt
to compute the effects due to a changing shape of the $p_T$ spectrum
caused by collective flow.  Distortions of the $p_T$ spectrum folded
into detection biases might
have the effect of artificially enhancing or suppressing
the $\eta^\prime$ contribution.  Additional contributions come
from dileptons produced in hadron-hadron collisions during the
expansion and cooling phase, which help to fill-in not only the
mass region between 2$m_{\pi}$ and $m_{\rho}$ but also the region
between the $\phi$ and the $J/\psi$ mesons \cite{peter}.

We should caution the reader that a big enhancement of $\eta'$
production would probably cause a suppression of direct production
of other mesons due to energy conservation.  For example, if
the only mesons produced were the $\eta'$ and the neutral and
charged pions, and if $\eta'/\pi^0$ was increased from 0.02 to 1,
then the total number of outcoming pions, including those from
$\eta'$ decay, would approximately double.  It would be a good
exercise to refit the abundances of all the mesons with this effect
taken into account.  Of course, the total number of mesons could
still increase, with the required energy coming from a decrease
in the average momentum of the particles.  This ties in with
the problem of distortion of the $p_T$ spectrum due to collective
flow.

Perhaps the most convincing demonstration of the return of the $\eta^\prime$
would be a direct measurement.  This might be possible for the two
photon decay mode, especially if the production cross section is
as strongly enhanced as we suggest.  It would be important to have
a simultaneous direct measurement of the $\eta$ since we expect an
enhancement there too.  In fact, some enhancement of the $\eta/\pi^0$
ratio in central S+Au collisions was indeed observed experimentally
by the WA80 experiment \cite{wa80}.  In minimum bias events the
ratio was measured to be 0.29$\pm$0.13, consistent with proton-proton
collisions.  In central collisions the ratio was measured to be
0.54$\pm$0.14.  Both are integrated ratios from $p_T$ = 0 to
1 GeV/c.  Since the branching ratio of $\eta'$ into $\eta$ is
about 65\%, an enhancement of $\eta'/\pi^0$ = 1 is close to being
ruled out (but recall the caveats about energy conservation and
$p_T$ distortion mentioned above).  An enhancement of $\eta'/\pi^0$ = 0.3
is more consistent with this data and more theoretically likely.

We should emphasize that unlike the case for the $\rho$ meson, and to a lesser
degree for the $\omega$ and $\phi$, the $\eta^\prime$ and the $\eta$ mesons
almost always decay after the surrounding hadronic matter has blown
apart. Therefore one cannot expect to directly see the effect of the
mass shift of the $\eta^\prime$ or the $\eta$ meson: the only effect will
be due to an enhanced production cross section.

\section*{Acknowledgments}

We gratefully acknowledge conversations with D. Dyakonov, H. Satz,
E. Shuryak, R. Thews, I. Tserruya, X.-N. Wang and I. Zahed.
We thank Jean Tran Thanh Van and the organizers of the XXX
Rencontres de Moriond where this work was initiated, and B. M\"uller,
K. Kinder-Geiger, J.-Y. Ollitrault and the ECT* in Trento, Italy
for hospitality during the workshop on QCD and Ultrarelativistic
Heavy Ion Collisions where this work was completed.
This work was supported by the U.S.
Department of Energy under grant numbers DE-FG02-87ER40328 and
DE-FG02-94ER40823 and by the German Research Ministry (BMFT)
under contract 06 BI 721.

\section*{Figure Captions}

\noindent
Figure 1:  Yield of low mass dileptons as measured by CERES \cite{ceres}.
Included in the plot are their assumed resonance contributions.
The heavy shaded area is the result of summing all these contributions,
including estimated uncertainties.

\noindent
Figure 2:  The two curves are the result of multiplying the assumed
$\eta'$ constribution in Figure 1 by factors of 16 and 50, and
adding the other contributions.

\end{document}